\documentclass[12pt,a4paper]{article}
\usepackage{amsmath}
\usepackage[space]{cite}
\usepackage[total={16cm,24cm},centering,includeheadfoot]{geometry}
\numberwithin{equation}{section}
\newcommand\preprint[1]{\begin{flushright}\normalfont{#1}
    \end{flushright}\vspace*{0ex}}
\renewcommand\title[1]{\thispagestyle{empty}\begin{center}
    \LARGE{#1}\end{center}\vspace{-1ex}\par
    }
\renewcommand\author[1]{\begin{center}\large{#1}\end{center}}
\newcommand\address[1]{\begin{center}
    \textit{#1}\end{center}\vspace{0ex}
    }

\renewcommand\thanks[1]{%
    \renewcommand{\thefootnote}{\fnsymbol{footnote}}\footnote{#1}}

\renewenvironment{abstract}{\par\bigskip
    \begin{center}\textbf{Abstract}\end{center}\par}{}

\newcommand\maketitlepage{\thispagestyle{empty}\clearpage
    \renewcommand{\thefootnote}{\arabic{footnote}}}

\newcommand{\rme}{\textrm{e}}
\newcommand{\rmi}{\textrm{i}}
\newcommand{\eps}{\varepsilon}

\newcommand{\const}{\textrm{const}\times}

\newcommand{\F}[4]{%
    {}\,{}_2F_1\biggl(%
    \genfrac{}{}{0pt}{}{#1\,,\:#2}{#3}
    \bigg\vert#4\biggr)}
\newcommand{\Fthreetwo}[6]{%
    \,{}_3F_2\biggl(%
    \genfrac{}{}{0pt}{}{#1\,,\:#2\,,\:#3}{#4\,,\:#5}
    \bigg\vert#6\biggr)}

\renewcommand{\Re}{\mathrm{Re}\,}

\begin{document}
%%%%%%%%%%%%%%%%%%%%%%%%%%%%%%%%%%%%%%%%%%%%%%%%%%%%%%%%%%%%%%%%%%%%
\preprint{DFF 429/11/05}
\title{
The role of orthogonal polynomials in the six-vertex model
and  its combinatorial applications\thanks{
Talk presented by F.C. at the
Short Program of the Centre de Recherches Math\'ematiques:
\textit{`Random Matrices, Random Processes and Integrable Systems'},
Montr\'eal,  June 20\textsuperscript{th}--July 8\textsuperscript{th}, 2005.}
}

\setcounter{footnote}{6}

\author{
F. Colomo and A. G. Pronko\thanks{On leave of absence from:
Saint Petersburg Department of V. A. Steklov
Mathematical Institute
of Russian Academy of Sciences,
Fontanka 27, 191023 Saint Petersburg, Russia.}
}

\setcounter{footnote}{1}

\address{
I.N.F.N., Sezione di Firenze,
and Dipartimento di Fisica, Universit\`a di Firenze,\\
Via G. Sansone 1, 50019 Sesto Fiorentino (FI), Italy}

\begin{abstract}
The Hankel determinant representations for the partition function
and  boundary correlation functions of the
six-vertex model with domain wall boundary conditions
are investigated by the methods of orthogonal polynomial theory.
For specific values of the parameters of the model,
corresponding to $1$-, $2$- and
$3$-enumerations of Alternating Sign Matrices (ASMs), these polynomials
specialize  to classical ones (Continuous Hahn, Meixner-Pollaczek, and
Continuous Dual Hahn, respectively). As a consequence,
a unified and simplified treatment of ASMs enumerations turns out
to be possible, leading also to some new results such as the
refined $3$-enumerations of ASMs. Furthermore,
the use of orthogonal polynomials allows us to express,
for generic values of the parameters of the model,
the partition function of the
(partially) \emph{inhomogeneous} model in terms of the one-point boundary
correlation functions of the \emph{homogeneous} one.
\end{abstract}

\maketitlepage
%\setcounter{page}{2}
%%%%%%%%%%%%%%%%%%%%%%%%%%%%%%%%%%%%%%%%%%%%%%%%%%%%%%%%%%%%%%%%%%%%%%%
\section{Introduction}

We shall consider here the six-vertex model
on a square lattice with domain wall
boundary conditions (DWBC).
The  model, in its inhomogeneous formulation, i.e., with
position dependent Boltzmann weights,
was originally proposed  in \cite{K-82}, within the theory of
correlation functions of quantum integrable models, in the framework
of the quantum inverse scattering method \cite{KBI-93}.
It was subsequently solved in \cite{I-87}, where a determinant formula for the
partition function was obtained and proven (see also \cite{ICK-92}).
Analogous determinant formulae have been given also
for the one- and two-point
boundary correlation functions \cite{BPZ-02,CP-05b}.
In its homogeneous version, the six-vertex model with DWBC
admits usual interpretation as a model of statistical mechanics
with fixed boundary conditions,
and it may be seen as a variation of the original model
with periodic boundary conditions \cite{LW-72,B-82}.
The model is known to be closely related with the problems
of enumeration of alternating sign matrices (ASMs) and domino tilings
(see book \cite{Br-99} for a nice review).
It should be mentioned that ASM enumerations in turn emerge,
via Razumov-Stroganov conjecture  \cite{RS-01,RS-04},
in the context of some quantum spin chains and loop models;
for recent works, see
for instance \cite{NRdG-05,DfZj-05,dGN-05} and references therein.

Till now, specific results for the six-vertex model with DWBC
at particular values of its parameters (obtained mainly in application
to ASM enumerations)
were derived from general results for the inhomogeneous version,
first specializing the parameters to the considered case,
and next performing the homogeneous limit. Each time, the homogeneous
limit was an hard task on its own right,
and a specific approach was devised  to work it out in each
single case \cite{Ku-96,Z-96b,S-02}. More recently, especially
in the context of Razumov-Stroganov conjecture, where the model
is considered at the so-called ice-point, the homogeneous limit of
the partition function is often performed only on a subset of the spectral
parameters, the remaining ones being reinterpreted as the variables
of some generating function encoding some peculiar properties of the model
\cite{S-02,DfZj-04}.

Here we consider a different approach, which turns
out to be rather convenient: we start directly from the homogeneous limit
representations, worked out once for all for the model with generic vertex
weights. Such `Hankel determinant' representations has been
derived in \cite{ICK-92} for the partition function, and in
\cite{BPZ-02} and \cite{CP-05b}, respectively, for the one- and two-point
boundary correlators. Following a rather standard procedure,
these quantities can then be expressed
in terms of orthogonal  polynomials. For
specific values of the vertex weights corresponding to the so-called
`ice-point', `free-fermion line' and `dual ice-point' of
the six-vertex model, these orthogonal polynomials specialize to
classical ones
(i.e., of hypergeometric type, or, equivalently, belonging to the Askey
scheme, see for example \cite{KS-98}), allowing the evaluation of
the partition function and boundary correlators
in closed form. These three cases
correspond  to the $1$-, $2$- and $3$-enumerations of ASMs,
respectively; a unified and
simplified treatment  of known ASMs is thus
provided, leading also to some
new results such as the refined $3$-enumerations of ASMs.

For generic values of the vertex weights, such orthogonal
polynomial representations
allow to express the two-point boundary correlator
in terms of the analogous one-point boundary correlator.
Such relationship can in turn be understood
as follows: the partition function
of the model  with two inhomogeneities may be written
in terms of the one point boundary correlator solely.
This simple result can be extended further: indeed here we
show that it is possible to
`get away' from the homogeneous limit and recover the structure of
the inhomogeneous partition function, thus generalizing previous formulae
of Stroganov \cite{S-02} to a generic number of  unfixed spectral parameters.
Specifically, we provide a simple  expression for the
partition function of the (partially) \emph{inhomogeneous} model
in terms of the one-point boundary  correlators of the \emph{homogeneous}
model; our result is valid for any choice
of the crossing parameter.  This somehow simplifies and
render explicit previous results discussed in \cite{DfZj-04},
where the  partition function of the inhomogeneous model (at ice-point)
was expressed in terms  of Shur functions.

%%%%%%%%%%%%%%%%%%%%%%%%%%%%%%%%%%%%%%%%%%%%%%%%%%%%%%%%%%%%%%%%%%%
\section{The six-vertex model with DWBC}\label{sec.6vm}

\begin{figure}[t]
%%%%%%%%%%%%%%%%%%%%%%%%%%%%%%%%%%%%%%%%%%%%%%%%%%%%%%
%%%%%%%%%% Figure 1 %%%%%%%%%%%%%%%%%%%%%%%%%%%%%%%%%%
%%%%%%%%%%%%%%%%%%%%%%%%%%%%%%%%%%%%%%%%%%%%%%%%%%%%%%
\unitlength=1mm
\begin{center}
\begin{picture}(124,30)
% ARROWS
% vertex a1
\put(7,20){\line(0,1){10}}
\put(7,22.5){\vector(0,-1){1}}
\put(7,27.5){\vector(0,-1){1}}
\put(2,25){\line(1,0){10}}
\put(4.5,25){\vector(-1,0){1}}
\put(9.5,25){\vector(-1,0){1}}
% vertex a2
\put(27,20){\line(0,1){10}}
\put(27,22.5){\vector(0,1){1}}
\put(27,27.5){\vector(0,1){1}}
\put(22,25){\line(1,0){10}}
\put(24.5,25){\vector(1,0){1}}
\put(29.5,25){\vector(1,0){1}}
% vertex b1
\put(52,20){\line(0,1){10}}
\put(52,22.5){\vector(0,-1){1}}
\put(52,27.5){\vector(0,-1){1}}
\put(47,25){\line(1,0){10}}
\put(49.5,25){\vector(1,0){1}}
\put(54.5,25){\vector(1,0){1}}
% vertex b2
\put(72,20){\line(0,1){10}}
\put(72,22.5){\vector(0,1){1}}
\put(72,27.5){\vector(0,1){1}}
\put(67,25){\line(1,0){10}}
\put(69.5,25){\vector(-1,0){1}}
\put(74.5,25){\vector(-1,0){1}}
% vertex c1
\put(97,20){\line(0,1){10}}
\put(97,22.5){\vector(0,1){1}}
\put(97,27.5){\vector(0,-1){1}}
\put(92,25){\line(1,0){10}}
\put(94.5,25){\vector(-1,0){1}}
\put(99.5,25){\vector(1,0){1}}
% vertex c2
\put(117,20){\line(0,1){10}}
\put(117,22.5){\vector(0,-1){1}}
\put(117,27.5){\vector(0,1){1}}
\put(112,25){\line(1,0){10}}
\put(114.5,25){\vector(1,0){1}}
\put(119.5,25){\vector(-1,0){1}}
% LINES
% vertex a1
\linethickness{0.5mm}
\put(7.25,10.25){\line(0,1){4.75}}
\put(7.5,10.25){\line(-1,0){5}}
\put(7.75,9.75){\line(0,-1){4.75}}
\put(7.5,9.75){\line(1,0){5}}
% vertex a2
\thinlines
\put(27,5){\line(0,1){10}}
\put(22,10){\line(1,0){10}}
% vertex b1
\linethickness{0.5mm}
\put(51.75,5){\line(0,1){10}}
\thinlines
\put(47,10){\line(1,0){10}}
% vertex b2
\thinlines
\put(72,5){\line(0,1){10}}
\linethickness{0.5mm}
\put(67,9.75){\line(1,0){10}}
% vertex c1
\linethickness{0.5mm}
\put(96.75,10){\line(0,1){5}}
\put(96.9,10){\line(-1,0){5}}
\thinlines
\put(97,9.75){\line(0,-1){5}}
\put(97,9.7){\line(1,0){4.7}}
% vertex c2
\put(117,10.25){\line(0,1){4.75}}
\put(117,10.25){\line(-1,0){5}}
\linethickness{0.5mm}
\put(117.25,10){\line(0,-1){4.75}}
\put(117,10){\line(1,0){5}}
% WEIGHTS
\put(5,-5){$w_1$}
\put(25,-5){$w_2$}
\put(50,-5){$w_3$}
\put(70,-5){$w_4$}
\put(95,-5){$w_5$}
\put(115,-5){$w_6$}
\end{picture}
\end{center}
\caption{The six allowed types of vertices
in terms of arrows (first row),
in terms of lines (second row), and their Boltzmann weights
(third row).}
\label{vertices}
\end{figure}
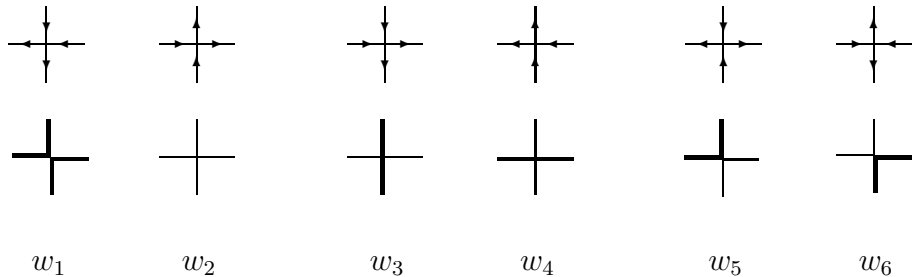

The six-vertex model, which was originally proposed as
a model of two-dimensional ice (hence the alternative
denomination: `square ice'),
is formulated on a square lattice
with arrows lying on edges, and
obeying the so-called `ice-rule', namely, the
only admitted configurations are such that there are always two
arrows pointing away from, and two arrows pointing into, each lattice
vertex.
An equivalent and  graphically simpler description
of the configurations of the model can be given
in terms of lines flowing through the vertices: for each arrow
pointing downward or to the left, draw a thick line on the
corresponding
link. The six possible vertex states
and the Boltzmann weights $w_i$
assigned to each vertex according
to its state $i$ ($i=1,\dots,6$) are shown in figure \ref{vertices}.
We shall presently restrict ourselves  to the homogeneous version
of the model, where the Boltzmann weights are site independent. We shall
however turn to the inhomogeneous version of the model
in the last section.

\begin{figure}[t]
%%%%%%%%%%%%%%%%%%%%%%%%%%%%%%%%%%%%%%%%%%%%%%%%%%%%%
%%%%%%%%%% figure 2 %%%%%%%%%%%%%%%%%%%%%%%%%%%%%%%%%
%%%%%%%%%%%%%%%%%%%%%%%%%%%%%%%%%%%%%%%%%%%%%%%%%%%%%
\unitlength=1mm
\begin{center}
\begin{picture}(30,30)
\put(0,5){\line(1,0){30}}
\put(5,5){\vector(-1,0){3.5}}
\put(10,5){\vector(-1,0){3.5}}
\put(15,5){\vector(-1,0){3.5}}
\put(20,5){\vector(-1,0){3.5}}
\put(25,5){\vector(-1,0){3.5}}
\put(25,5){\vector(1,0){3.5}}
\put(0,10){\line(1,0){30}}
\put(5,10){\vector(-1,0){3.5}}
\put(10,10){\vector(-1,0){3.5}}
\put(10,10){\vector(1,0){3.5}}
\put(15,10){\vector(1,0){3.5}}
\put(20,10){\vector(1,0){3.5}}
\put(25,10){\vector(1,0){3.5}}
\put(0,15){\line(1,0){30}}
\put(5,15){\vector(-1,0){3.5}}
\put(10,15){\vector(-1,0){3.5}}
\put(15,15){\vector(-1,0){3.5}}
\put(20,15){\vector(-1,0){3.5}}
\put(20,15){\vector(1,0){3.5}}
\put(25,15){\vector(1,0){3.5}}
\put(0,20){\line(1,0){30}}
\put(5,20){\vector(-1,0){3.5}}
\put(5,20){\vector(1,0){3.5}}
\put(15,20){\vector(-1,0){3.5}}
\put(15,20){\vector(1,0){3.5}}
\put(20,20){\vector(1,0){3.5}}
\put(25,20){\vector(1,0){3.5}}
\put(0,25){\line(1,0){30}}
\put(5,25){\vector(-1,0){3.5}}
\put(10,25){\vector(-1,0){3.5}}
\put(10,25){\vector(1,0){3.5}}
\put(15,25){\vector(1,0){3.5}}
\put(20,25){\vector(1,0){3.5}}
\put(25,25){\vector(1,0){3.5}}
\put(5,0){\line(0,1){30}}
\put(5.05,0){\vector(0,1){3.5}}
\put(5.05,5){\vector(0,1){3.5}}
\put(5.05,10){\vector(0,1){3.5}}
\put(5.05,15){\vector(0,1){3.5}}
\put(5.05,25){\vector(0,-1){3.5}}
\put(5.05,30){\vector(0,-1){3.5}}
\put(10,0){\line(0,1){30}}
\put(10.05,0){\vector(0,1){3.5}}
\put(10.05,5){\vector(0,1){3.5}}
\put(10.05,15){\vector(0,-1){3.5}}
\put(10.05,20){\vector(0,-1){3.5}}
\put(10.05,20){\vector(0,1){3.5}}
\put(10.05,30){\vector(0,-1){3.5}}
\put(15,0){\line(0,1){30}}
\put(15.05,0){\vector(0,1){3.5}}
\put(15.05,5){\vector(0,1){3.5}}
\put(15.05,10){\vector(0,1){3.5}}
\put(15.05,15){\vector(0,1){3.5}}
\put(15.05,25){\vector(0,-1){3.5}}
\put(15.05,30){\vector(0,-1){3.5}}
\put(20,0){\line(0,1){30}}
\put(20.05,0){\vector(0,1){3.5}}
\put(20.05,5){\vector(0,1){3.5}}
\put(20.05,10){\vector(0,1){3.5}}
\put(20.05,20){\vector(0,-1){3.5}}
\put(20.05,25){\vector(0,-1){3.5}}
\put(20.05,30){\vector(0,-1){3.5}}
\put(25,0){\line(0,1){30}}
\put(25.05,0){\vector(0,1){3.5}}
\put(25.05,10){\vector(0,-1){3.5}}
\put(25.05,15){\vector(0,-1){3.5}}
\put(25.05,20){\vector(0,-1){3.5}}
\put(25.05,25){\vector(0,-1){3.5}}
\put(25.05,30){\vector(0,-1){3.5}}
\put(12.5,-5){(a)}
\end{picture}
\qquad
\begin{picture}(30,30)
\put(0,5){\line(1,0){30}}
\put(0,10){\line(1,0){30}}
\put(0,15){\line(1,0){30}}
\put(0,20){\line(1,0){30}}
\put(0,25){\line(1,0){30}}
\put(5,0){\line(0,1){30}}
\put(10,0){\line(0,1){30}}
\put(15,0){\line(0,1){30}}
\put(20,0){\line(0,1){30}}
\put(25,0){\line(0,1){30}}
\linethickness{0.5mm}
\put(0,5){\line(1,0){25.25}}
\put(25,30){\line(0,-1){25.25}}
\put(0,10){\line(1,0){10.25}}
\put(20,30){\line(0,-1){15.25}}
\put(20.25,14.75){\line(-1,0){10}}
\put(10.25,9.75){\line(0,1){5.25}}
\put(0,15.25){\line(1,0){10}}
\put(15,30){\line(0,-1){10.25}}
\put(15,20){\line(-1,0){5.5}}
\put(9.75,20){\line(0,-1){5}}
\put(0,20){\line(1,0){5.25}}
\put(10,30){\line(0,-1){5.25}}
\put(10.25,24.75){\line(-1,0){5}}
\put(5.25,25){\line(0,-1){5.25}}
\put(0,25.25){\line(1,0){5}}
\put(4.75,30){\line(0,-1){5}}
\put(12.5,-5){(b)}
\end{picture}
\end{center}
\caption{One of the possible configurations of the model with DWBC, in
the case$N=5$:
(a) in terms of arrows; (b) in terms of lines.}
\label{dwbcgrid}
\end{figure}
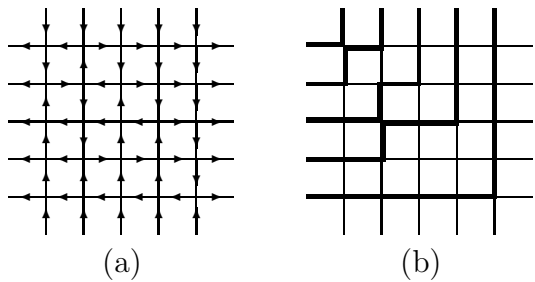

The DWBC are imposed on the $N\times N$ square lattice by fixing
the  direction of  all arrows on the boundaries in a specific way.
Namely, the vertical
arrows on the top and bottom
of the lattice point inward, while the horizontal arrows on the left
and right sides point outward. Equivalently, a generic configuration
of the model with  DWBC  can
be depicted by $N$ lines flowing from the upper boundary to the left one.
This line picture (besides taking into account the `ice-rule'
in an automated way) is intuitively closer to ASMs
recalled in the next section. A possible state of the model both in terms
of arrows and of lines is shown in figure \ref{dwbcgrid}.

The partition function is defined, as usual, as a
sum over all possible arrow configurations, compatible with
the imposed DWBC, each configuration being assigned its Boltzmann weight,
given as the product of all the corresponding vertex weights,
\begin{equation}%\label{}
Z_N=\sum_{\substack{\text{arrow configurations}\\
\text{with DWBC}}}^{}\
\prod_{i=1}^{6}w_i^{n_i}\;.
\end{equation}
Here $n_i$ denotes the number of vertices in the state $i$
in each arrow configuration ($n_1+\dots+n_6=N^2$).

The six-vertex model with DWBC can be considered,
with no loss of generality,
with its weights invariant under the simultaneous reversal
of all arrows,
\begin{equation}%\label{}
w_1=w_2=:a,\qquad
w_3=w_4=:b,\qquad
w_5=w_6=:c.
\end{equation}
Under different choices of Boltzmann weights
the six-vertex model exhibits different  behaviours,
according to the value of the parameter $\Delta$, defined as
\begin{equation}
\Delta=\frac{a^2+b^2-c^2}{2ab}.
\end{equation}
It is well known that
there are three physical regions or phases for the six-vertex model:
the ferroelectric phase, $\Delta>1$;
the anti-ferroelectric phase, $\Delta<-1$;
and, the disordered phase,
$-1<\Delta<1$. Here we restrict ourselves to
the disordered phase, where the Boltzmann weights are conveniently
parameterized as
\begin{equation} \label{sin}
a=\sin(\lambda+\eta),\qquad
b=\sin(\lambda-\eta),\qquad
c=\sin 2\eta.
\end{equation}
With this choice one has $\Delta=\cos 2\eta$. The parameter $\lambda$
is the so-called spectral parameter
and $\eta$ is the crossing parameter.
The physical requirement of positive Boltzmann weights, in the disordered
regime, restricts the values of the  crossing and  spectral parameters
to $0<\eta<\pi/2$ and $\eta<\lambda<\pi-\eta$.

An exact representation for the partition function
(for generic weights, even complex) was
obtained in \cite{ICK-92}.
When the weights are parameterized according to \eqref{sin}
such representation reads
\begin{equation}\label{ZNhom}
Z_N=\frac{[\sin(\lambda-\eta)\sin(\lambda+\eta)]^{N^2}}
{\prod_{n=1}^{N-1}(n!)^2}\;
{\det}_N^{} \varPhi
\end{equation}
where $\varPhi$ is an $N\times N$ matrix with entries
\begin{equation}\label{varphi}
\varPhi_{jk}=\partial_{\lambda}^{j+k}
\varphi(\lambda,\eta),\qquad
\varphi(\lambda,\eta)=
\frac{\sin(2\eta)}{\sin(\lambda-\eta)\sin(\lambda+\eta)}.
\end{equation}
Here and in the following we use the convention that indices
of $N\times N$ matrices run
over the values $j,k=0,\dots,N-1$.

This formula for the partition function has been obtained
as the homogeneous limit of a  more
general formula for the inhomogeneous
six-vertex model with DWBC. The inhomogeneous model,
with site-dependent weights, is defined by introducing
two sets of spectral parameters $\{\lambda_{\alpha}\}_{\alpha =1}^N$ and
$\{\nu_{\beta}\}_{\beta =1}^N$, such that the weights
of the vertex lying at the
intersection of the $\alpha $-th column with the $\beta $-th row depends on
$\lambda_\alpha -\nu_\beta $ rather than simply on  $\lambda$, still
through formulae \eqref{sin}. The inhomogeneous model
can be fruitfully investigated through  the
Quantum Inverse Scattering Method,  see papers \cite{K-82,I-87,ICK-92,dGK-01}
and book \cite{KBI-93} for details. As a  result, the partition
function of the inhomogeneous model is represented in terms of
certain determinant
formula which, however, requires some effort in the study of its
homogeneous limit, $\nu_\beta \to 0$ and $\lambda_\alpha \to\lambda$, since
in this limit the determinant possesses $N^2-N$ zeros
that are cancelled by the same number of singularities coming
from the pre-factor. A recipe for taking such a
limit was explained in \cite{ICK-92}
where formula \eqref{ZNhom} was originally obtained.
Subsequently, formula \eqref{ZNhom} was used in papers
\cite{KZj-00,Zj-00} to investigate the thermodynamic limit, $N\to\infty$,
of the partition function. In these  studies the
Hankel structure of the determinant appearing in \eqref{ZNhom}, a
natural outcome of the  homogeneous limit procedure, was exploited
through its relation with   the  Toda chain differential equation
and with matrix models.

In addition to the partition function, in the following we shall discuss
one- and two-point boundary correlation functions as well.
In general,  two kinds of one-point correlation functions
can be considered in the six-vertex model: the first one (`polarization')
is the probability to find an arrow on a given edge in a particular
state,
while the second one is the probability to find a given vertex in
some state $i$.
If one restricts to edges or vertices adjacent to the boundary,
then such correlation functions are called boundary correlation functions.
Following the notations of paper \cite{BPZ-02}, where these boundary
correlation functions were studied, let $G_N^{(r)}$ denote the probability
that an arrow on the first row  and between the $r$-th and  $(r+1)$-th
columns (enumerated from the right) points left
(or, in the line language,
that there is a thick line on this edge), and let $H_N^{(r)}$ denote
the probability that the first vertex in the $r$-th column (counted from
the right) is in the
state $i=5$ (or that the thick line flows from the top to the left),
see Figs.~\ref{vertices} and \ref{dwbcgrid}.
The first correlation function, $G_N^{(r)}$, is, in fact, the
boundary polarization, whose interpretation is more direct from
a physical point of view, while
the second one,  $H_N^{(r)}$, is closely related to the refined
enumerations of ASMs. It is easy to see that,
due to DWBC, the two correlation functions are related to each other as follows
\begin{equation}\label{GviaH}
G_N^{(r)}= H_N^{(r)} + H_N^{(r-1)} +\cdots + H_N^{(1)}.
\end{equation}

In \cite{BPZ-02} both correlation functions were computed using Quantum
Inverse Scattering Method for the inhomogeneous six-vertex model.
In the homogeneous limit, which is the situation we are interested
in here, determinant formulae generalizing \eqref{ZNhom} were
found for these correlation functions. For instance, for $H_N^{(r)}$,
the following expression was derived
\begin{equation}\label{Hhom}
H_N^{(r)}=\frac{(N-1)!\,\sin(2\eta)}
{\big[\sin(\lambda+\eta)\big]^r\big[\sin(\lambda-\eta)\big]^{N-r+1}}\;
\frac{{\det}_N \varPsi}{{\det}_N \varPhi}
\end{equation}
where the matrix $\varPsi$ differs from the matrix $\varPhi$,
equation \eqref{varphi}, just by the elements of the last column
\begin{equation}%\label{}
\varPsi_{j ,k}=\varPhi_{j ,k},\quad k=0,\dots,N-2;\qquad
\varPsi_{j ,N}=\partial_\eps^{j}
\frac{(\sin\eps)^{N-r}[\sin(\eps-2\eta)]^{r-1}}
{[\sin(\eps+\lambda-\eta)]^{N-1}}\bigg|_{\eps=0}.
\end{equation}
A similar expression is valid for $G_N^{(r)}$ as well.
In what follows we shall focus on $H_N^{(r)}$; the results for
$G_N^{(r)}$ will follow immediately from relation \eqref{GviaH}.
{}From the DWBC it immediately follows that $G_N^{(N)}=1$, and
therefore, from \eqref{GviaH},  correlation function $H_N^{(r)}$
satisfies
\begin{equation}\label{normcond}
\sum_{r=1}^N \ H_N^{(r)} = 1 \,.
\end{equation}
In the following this normalization condition
will be used in application to the generating function of
$H_N^{(r)}$, defined as
\begin{equation}\label{Hnu}
H_N(u):= \sum_{r=1}^{N} H_N^{(N-r+1)} u^{r-1}.
\end{equation}

Let us finally recall the two-point boundary correlation function.
Again several different definitions are sensible. We shall focus here,
for the sake of simplicity and definiteness, on a specific
two-point generalization of the one-point
correlation function $H_N^{(r)}$, describing
the probability of finding vertices of type $i=5$
on the opposite, top and bottom, boundaries. More precisely, we define
$H^{(r_1,r_2)}_N$ as the probability of finding vertices of type
$i=5$  both  at the  $r_1$-th position
of the first row and  at the $r_2$-th position of the last row
($r_1$ and $r_2$ are counted from the right).

In \cite{CP-05b} such correlation function was  computed using Quantum
Inverse Scattering Method for the inhomogeneous six-vertex model.
In the homogeneous limit, which is the situation we are interested
in here, the following determinant formula was derived:
\begin{multline}\label{H2hom}
H_N^{(r_1,r_2)}=
\frac{(N-1)!\,(N-2)!\,\sin^2(2\eta)}
{\big[\sin(\lambda+\eta)\big]^{N+r_1-r_2+1}
\big[\sin(\lambda-\eta)\big]^{N+r_2-r_1+1}
{\det}_N \varPhi}
\\  \times
\left[
{\det}_N \Big(\varPhi_{j ,k}\Big|
\partial_{\eps_2}^{j}\Big|\partial_{\eps_1}^{j}
\Big)_{0\leq j  \leq N-1,0\leq k\leq N-3}\;
h_N^{(r_1,r_2)}(\eps_1,\eps_2)\right]\bigg|_{\eps_1=\eps_2=0}.
\end{multline}
Here the function $h_N^{(r_1,r_2)}(\eps_1,\eps_2)$ is defined as follows:
\begin{equation}\label{h2}
h_N^{(r_1,r_2)}(\eps_1,\eps_2)=
\frac{(\sin\eps_1)^{N-r_1}[\sin(\eps_1-2\eta)]^{r_1-1}
(\sin\eps_2)^{N-r_2}[\sin(\eps_2+2\eta)]^{r_2-1}}
{\sin(\eps_2-\eps_1+2\eta)[\sin(\eps_1+\lambda-\eta)]^{N-2}
[\sin(\eps_2+\lambda+\eta)]^{N-2}}.
\end{equation}
The matrix $\varPhi$ is defined in \eqref{varphi}.

It is obvious from the definitions of $H_N^{(r)}$ and
$H_N^{(r_1,r_2)}$, that
\begin{equation}\label{Htwonorm}
\sum_{r_1=1}^N H_N^{(r_1,r)} =
\sum_{r_2=1}^N H_N^{(r,r_2)} =
H_N^{(r)}\,, \qquad \qquad\sum_{r_1,r_2=1}^N H_N^{(r_1,r_2)} =1\,.
\end{equation}
In dealing with the two-point boundary function
it will be convenient to use the corresponding generating function
\begin{equation}\label{Hnuv}
H_N(u,v):= \sum_{r=1,s=1}^{N} H_N^{(N-r+1,s)} u^{r-1} v^{s-1}.
\end{equation}
Note that \eqref{Htwonorm} simply implies
$H_N(1,v)=H_N(v,1)=H_N(v)$ where $H_N(v)$ is defined by \eqref{Hnu};
we also have $H_N(1,1)=1$.

%The latter  relation can also be viewed as a normalization
%condition for  the generating function
%\begin{equation}\label{Hnuv}
%H_N(u,v):= \sum_{r=1,s=1}^{N} H_N^{(N-r+1,s)} u^{r-1} v^{s-1}
%\end{equation}
%of  the two-point boundary correlation function.

%%%%%%%%%%%%%%%%%%%%%%%%%%%%%%%%%%%%%%%%%%%%%%%%%%%%%%%%%%%%%%%%%%%
\section{Orthogonal polynomial representation}\label{sec.orthog}

\subsection{Preliminaries}

In this section the previously reviewed determinant representations
for the partition function and the one- and two-point
boundary correlation functions  will be analysed by making
use of the orthogonal polynomial theory, along the lines proposed
in paper \cite{CP-05a,CP-05b}. We start with recalling some very standard
and well-known facts from the general theory.

Let $\{P_n(x)\}_{n=0}^\infty$ be a set of polynomials, with
non-vanishing leading coefficient
\begin{equation}%\label{}
P_n(x)=\kappa_n x^n + \dots,\qquad \kappa_n\ne0,
\end{equation}
and orthogonal on the real axis with respect to some
weight $\mu(x)$,
\begin{equation}\label{ortho}
\int_{-\infty}^{\infty} P_{n}(x)\, P_{m}(x)\, \mu(x)\, \mathrm{d} x
=h_{n} \delta_{nm}\,.
\end{equation}
Let $c_n$ denote $n$-th  moment of the weight $\mu(x)$, i.e.
\begin{equation}%\label{}
c_n =\int_{-\infty}^{\infty} x^n \mu(x) \mathrm{d} x,\qquad
n=0,1,\ldots
\end{equation}
and let us consider the $(n+1)\times (n+1)$ determinant
\begin{equation}%\label{}
D_n =
\begin{vmatrix}
c_0&c_1& \dots &c_n \\
c_1& c_2&  \dots & c_{n+1}\\
\hdotsfor{4}  \\
c_n & c_{n+1} & \dots & c_{2n}
\end{vmatrix}.
\end{equation}
Using the orthogonality condition \eqref{ortho} and
well-known properties of determinants, one can easily
derive the following formula
\begin{equation}\label{hankel}
D_n=\prod_{k=0}^n \frac{h_k}{\kappa_k^2}.
\end{equation}
This formula can be used for computation of Hankel determinants, provided
the orthogonal polynomials $\{P_n(x)\}_{n=0}^\infty$ are known.
On the other hand, the polynomials $\{P_n(x)\}_{n=0}^\infty$
can in turn be expressed
as determinants. For later use let us introduce the notation
\begin{equation}%\label{}
D_n^{(k)}(x_1,\dots,x_k) =
\begin{vmatrix}
c_0&c_1& \dots &c_{n-k}  & 1 & 1 & \dots & 1 \\
c_1&c_2& \dots &c_{n-k+1}& x_1 & x_2 &\dots &x_k \\
\hdotsfor{8}  \\
c_n& c_{n+1}&\dots & c_{2n-k} & x_1^n & x_2^n &\dots &x_k^n
\end{vmatrix}
\end{equation}
so that $D_n^{(0)}= D_n$, and $D_{n-1}^{(n)}(x_1, \dots,x_n) =
\varDelta(x_1,\dots,x_n)$, where $\varDelta(x_1,\dots,x_n)$ denotes
the Vandermonde determinant of $n$ variables.
For the polynomials one can find that
\begin{equation}\label{Px}
P_n(x)=\frac{\kappa_n}{ D_{n-1}} D_n^{(1)}(x).
\end{equation}
For a proof, see for example, \cite{S-75}.

Relation \eqref{Px} can be read off inversely thus giving
an expression for the determinant $ D_n^{(1)}(x)$
in terms of the polynomials $P_n(x)$. Taking into account that
(see \eqref{hankel})
\begin{equation}\label{kh}
\frac{h_n}{\kappa_n^2}= \frac{ D_n}{ D_{n-1}}
\end{equation}
we can write
\begin{equation}\label{D1}
\frac{ D_n^{(1)}(x)}{ D_n}= \frac{\kappa_n}{h_n}\; P_n(x).
\end{equation}

Consider now the case of $ D_n^{(2)}(x_1,x_2)$.
It is clear that the term of the highest powers on both
$x_1$ and $x_2$ is just
$ D_{n-2}(x_2^{n}x_1^{n-1}-x_1^{n} x_2^{n-1})$; following \cite{S-75},
it can be shown that
\begin{equation}%\label{}
 D_n^{(2)}(x_1,x_2)=
\frac{ D_{n-2}}{\kappa_n\kappa_{n-1}}
\big[P_{n-1}(x_1)P_n(x_2)-P_{n}(x_1)P_{n-1}(x_2)\big].
\end{equation}
Again using \eqref{kh}, we write
\begin{align}\label{D2}
\frac{ D_n^{(2)}(x_1,x_2)}
{ D_n}
&= \frac{\kappa_n\kappa_{n-1}}{h_nh_{n-1}}\;
\big[P_{n-1}(x_1)P_n(x_2)-P_{n}(x_1)P_{n-1}(x_2)\big]
\notag\\
& = \frac{\kappa_n\kappa_{n-1}}{h_nh_{n-1}}\;
\begin{vmatrix}
P_{n-1}(x_1) & P_{n-1}(x_2) \\
P_{n}(x_1) & P_{n}(x_2)
\end{vmatrix}.
\end{align}
This formula can be easily extended to the general case of
$ D_n^{(k)}(x_1,\dots,x_k)$.
We shall now use all these formulae to provide completely general
`orthogonal polynomial representations' for the partition function
and the one- and two-point boundary correlation functions.

\subsection{The partition function}

First we note, following papers \cite{Zj-00,CP-05a}, that the
matrix $\varPhi$ entering the expressions for the
homogenous model partition function and the boundary correlation functions
can be related with orthogonal
polynomials using the integral representation
\begin{equation}\label{int}
\frac{\sin(2\eta)}{\sin(\lambda-\eta)\sin(\lambda+\eta)} =
\int_{-\infty}^{\infty} \rme^{x(\lambda-\pi/2)}
\frac{\sinh(\eta x)}{\sinh(\pi x/2)}\; \mathrm{d}x.
\end{equation}
This formula is valid if $0<\eta<\pi/2$ and $\eta<\lambda<\pi-\eta$;
these values of $\lambda$ and $\eta$ correspond
to the so-called disordered regime of the six-vertex model.
When considering other regime of the model, the measure
of the corresponding polynomials becomes discrete, see \cite{Zj-00},
but the present procedure may nevertheless be considered, modulo trivial
modifications. Indeed, it can be easily seen that our results below
do not depend on the particular choice of the regime, and
can be extended to other regimes simply using the proper analytical
continuation in the parameters $\lambda$ and $\eta$.

Formula \eqref{int} implies that we have to deal with
the set of polynomials which are orthogonal with respect to
the following weight function
\begin{equation}
\label{meas}
\mu(x)=\mu(x;\lambda,\eta)=
\rme^{x(\lambda-\pi/2)}
\frac{\sinh(\eta x)}{\sinh(\pi x/2)}.
\end{equation}
The corresponding polynomials $P_n(x)=P_n(x;\lambda,\eta)$
also depend on $\lambda$ and $\eta$ which are to be considered
as parameters. In what follows we shall often omit the dependence
on $\lambda$ and $\eta$ where possible. Let us mention the following
useful property of these polynomials
\begin{equation}\label{cross}
P_n(x;\lambda,\eta)=(-1)^n\, P_n(-x;\pi-\lambda,\eta).
\end{equation}
This property can be easily established in virtue of formula
\eqref{Px}. It is to be mentioned also that both the leading coefficient
$\kappa_n=\kappa_n(\lambda,\eta)$ and the normalization constant
$h_n=h_n(\lambda,\eta)$ are invariant under the substitution
$\lambda\to\pi-\lambda$.
The transformation $\lambda\to\pi-\lambda$ is related to the
crossing symmetry of the six-vertex model, and the previous property
will have useful consequences in the discussion of the  one- and two-point
boundary correlation functions, discussed below.

Let us focus now on the partition function. The differential operator
in the Hankel matrix entering representation  \eqref{ZNhom}
pulls down powers of the integration variable,
and the matrix  itself is immediately rewritten as the
matrix of moments of the measure  \eqref{meas}.
Due to \eqref{hankel}, we readily get
\begin{equation}\label{detZ}
Z_N=\left(\frac{\sin 2\eta}{\varphi}\right)^{N^2}\,
\prod_{n=0}^{N-1}\frac{h_n}{(n!)^2 \kappa_n^2}
\end{equation}
where $\varphi=\varphi(\lambda,\eta)$ is exactly the function
defining entries of the matrix $\varPhi$, see \eqref{varphi}.

We just want to conclude with a simple remark:
whenever the values of the parameters $\lambda$ and $\eta$ are such
that the corresponding orthogonal polynomials $P_n(x;\lambda,\eta)$
happen to belong to the Askey scheme,
the evaluation of the partition function in closed form reduces
to an elementary computation.
As it will be shown in the following, this simple observation allows for
a direct and straightforward evaluation of ASM enumerations.

\subsection{The one-point boundary correlation function}

Let us now turn to the one-point boundary correlator. Starting
from its determinant representation, formula \eqref{Hhom},
and recalling relation \eqref{D1}, we readily write:
\begin{multline}\label{HNlong}
H_N^{(r)}(\lambda,\eta)=\frac{(N-1)!\,\sin(2\eta)}
{\big[\sin(\lambda+\eta)\big]^r\big[\sin(\lambda-\eta)\big]^{N-r+1}}\;
\frac{\kappa_{N-1}(\lambda,\eta)}{h_{N-1}(\lambda,\eta)}
\\ \times
P_{N-1}(\partial_\eps;\lambda,\eta)
\frac{(\sin\eps)^{N-r}[\sin(\eps-2\eta)]^{r-1}}
{[\sin(\eps+\lambda-\eta)]^{N-1}}\bigg|_{\eps=0}.
\end{multline}
This representation does not seem particularly appealing at this level,
but it is worth noticing that, when  the values of $\lambda$, $\eta$
are such that  $P_{N-1}(x,\lambda, \eta)$ reduces to a
classical polynomial,
the previous correlation function may be evaluated exactly, in closed form.
The use of the properties of the polynomial entering the representation is
a crucial ingredient in performing such computation.
This will be shown in detail later, but now we are interested
instead in obtaining another equivalent representation for $H_N^{(r)}$,
by making use of the crossing symmetry of the six-vertex model. The
use  of these two equivalent representations for the one-point boundary
correlation function $H_N^{(r)}$ will then allow us to express
the two-point boundary correlation function $H_N^{(r_1,r_2)}$ in terms of
one-point boundary correlation functions.

We  recall that
the crossing symmetry is the symmetry of the vertex weights
under reflection with respect to the vertical axis, and simultaneous
interchange of the functions $a$ and $b$, which is equivalent
to setting $\lambda\to\pi-\lambda$. Since the lattice with DWBC is
invariant under the reflection with respect to
the vertical axis (modulo reversal of all arrows on the horizontal edges),
the crossing symmetry  implies that the
following relation holds
\begin{equation}\label{crossym}
H_N^{(r)}(\lambda,\eta)=H_N^{(N-r+1)}(\pi-\lambda,\eta).
\end{equation}

Consider expression \eqref{Hhom} for the one-point function
$H_N^{(r)}$. Due to \eqref{D1} we can rewrite it as
\begin{multline}%\lable{HNlong}
H_N^{(r)}(\lambda,\eta)=\frac{(N-1)!\,\sin(2\eta)}
{\big[\sin(\lambda+\eta)\big]^r\big[\sin(\lambda-\eta)\big]^{N-r+1}}\;
\frac{\kappa_{N-1}(\lambda,\eta)}{h_{N-1}(\lambda,\eta)}
\\ \times
P_{N-1}(\partial_\eps;\lambda,\eta)
\frac{(\sin\eps)^{N-r}[\sin(\eps-2\eta)]^{r-1}}
{[\sin(\eps+\lambda-\eta)]^{N-1}}\bigg|_{\eps=0}.
\end{multline}
Taking into account \eqref{cross} and the properties of the
leading coefficient $\kappa_n(\lambda,\eta)$
and the normalization constant $h_n(\lambda,\eta)$ mentioned above,
it can be easily seen that from \eqref{HNlong} and \eqref{crossym}
the following expression is valid as well
\begin{multline}\label{HNlong2}
H_N^{(r)}(\lambda,\eta)=\frac{(N-1)!\,\sin(2\eta)}
{\big[\sin(\lambda+\eta)\big]^r\big[\sin(\lambda-\eta)\big]^{N-r+1}}\;
\frac{\kappa_{N-1}(\lambda,\eta)}{h_{N-1}(\lambda,\eta)}
\\ \times
P_{N-1}(\partial_\eps;\lambda,\eta)
\frac{(\sin\eps)^{r-1}[\sin(\eps+2\eta)]^{N-r}}
{[\sin(\eps+\lambda+\eta)]^{N-1}}\bigg|_{\eps=0}.
\end{multline}
Note that this expression means simply that the limit $\eps\to 0$ in
\eqref{HNlong} can be changed into $\eps\to 2\eta$ without altering
the result.

Now, these two equivalent representations,
\eqref{HNlong} and \eqref{HNlong2}, can be used in the study of
the two-point
correlation function $H_N^{(r_1,r_2)}$ given by
expression  \eqref{H2hom}, which certainly involves
similar structures.
Before turning to this analysis, let us put the above formulae
for the one-point function in a more compact and convenient
notations.

We define the functions
\begin{equation}%\label{}
\omega(\epsilon)=\frac{\sin(\lambda+\eta)}{\sin(\lambda-\eta)}\,
\frac{\sin\eps}{\sin(\eps-2\eta)},\qquad
\varrho(\epsilon)=\frac{\sin(\lambda-\eta)}{\sin(2\eta)}\,
\frac{\sin(\eps-2\eta)}{\sin(\eps+\lambda-\eta)},
\end{equation}
which are related to each other as
\begin{equation}\label{go}
\varrho(\eps)=\frac{1}{\omega(\eps)-1}.
\end{equation}
Also we define
\begin{equation}%\label{}
\tilde\omega(\epsilon)=\frac{\sin(\lambda-\eta)}{\sin(\lambda+\eta)}\,
\frac{\sin\eps}{\sin(\eps+2\eta)},\qquad
\tilde\varrho(\epsilon)=\frac{\sin(\lambda+\eta)}{\sin(2\eta)}\,
\frac{\sin(\eps+2\eta)}{\sin(\eps+\lambda+\eta)};
\end{equation}
which are in turn related to each other as
\begin{equation}\label{tgto}
\tilde\varrho(\eps)=\frac{1}{1-\tilde\omega(\eps)}.
\end{equation}
Note, that the functions with tildes are introduced such that
\begin{equation}%\label{}
\tilde\omega(\eps;\lambda,\eta)=\omega(-\eps;\pi-\lambda,\eta),\qquad
\tilde\varrho(\eps;\lambda,\eta)=-\varrho(-\eps;\pi-\lambda,\eta)
\end{equation}
in accordance with the crossing symmetry considerations made above.
Additionally, let us denote
\begin{equation}%\label{}
K_{N-1}(x)=(N-1)!\,\varphi^N\, \frac{\kappa_{N-1}}{h_{N-1}},
P_{N-1}(x)
\end{equation}
where $\varphi=\varphi(\lambda,\eta)$ is exactly the function
defining entries of the matrix $\varPhi$, see \eqref{varphi}.
In these notations formulae \eqref{HNlong} and \eqref{HNlong2}
for the correlation function $H_N^{(r)}$ read
\begin{equation}%\label{}
H_N^{(r)}=K_{N-1}(\partial_\eps)
\big[\omega(\eps)\big]^{N-r}[\varrho(\eps)]^{N-1}\Big|_{\eps=0}
\end{equation}
and
\begin{equation}%\label{}
H_N^{(r)}
= K_{N-1}(\partial_\eps)
\big[\tilde\omega(\eps)\big]^{r-1}
[\tilde\varrho(\eps)]^{N-1}\Big|_{\eps=0},
\end{equation}
respectively.

\subsection{The two-point boundary correlation function}

Let us now consider  the two-point correlation function
$H_N^{(r_1,r_2)}$, which is given by  formula \eqref{H2hom}.
Obviously, function \eqref{h2} contains all the
structures introduced above apart from the factor
$\sin(\eps_2-\eps_1+2\eta)$ standing in the denominator there.
However, using the identity
\begin{equation}%\label{}
\sin(2\eta) \sin(\eps_2-\eps_1+2\eta)=
\sin\eps_1 \sin\eps_2-
\sin(\eps_1-2\eta)\sin(\eps_2+2\eta)
\end{equation}
it can be easily seen that
\begin{equation}%\label{}
\frac{\sin(\eps_1+\lambda-\eta)
\sin(\eps_2+\lambda+\eta)}{\sin(\eps_2-\eps_1+2\eta)}
=\frac{1}{\varphi \varrho(\eps_1) \tilde \varrho(\eps_2)}\;
\frac{1}{\omega(\eps_1)\tilde\omega(\eps_2)-1}.
\end{equation}
Thus, taking into account formula \eqref{D2}
we can write the two-point correlation function in the form
\begin{multline}\label{H2nice}
H_N^{(r_1,r_2)}=
\left[K_{N-1}(\partial_{\eps_1})K_{N-2}(\partial_{\eps_2})
-K_{N-2}(\partial_{\eps_1})K_{N-1}(\partial_{\eps_2})\right]
\\ \times
\frac{\big[\omega(\eps_1)\big]^{N-r_1}[\varrho(\eps_1)]^{N-2}
\big[\tilde\omega(\eps_2)\big]^{N-r_2}[\tilde\varrho(\eps_2)]^{N-2}}
{\omega(\eps_1)\tilde\omega(\eps_2)-1}\bigg|_{\eps_1=0,\eps_2=0}.
\end{multline}

Taking into account that $\omega(\eps),\tilde \omega(\eps)\to 0$
as $\eps\to 0$
we can expand the denominator in \eqref{H2nice} in power series
and it can be easily seen that only the first few terms
(actually not more than $N$) of this expansion will contribute.
As a result, in virtue
of relations \eqref{go} and \eqref{tgto}, we arrive
to the following expression in terms of the one-point functions
\begin{multline}\label{H=HH}
H_N^{(r_1,r_2)}=
\sum_{j=1}^{N}
\Big(
H_N^{(r_1-j+1)} H_{N-1}^{(N-r_2+j)}
-H_N^{(r_1-j)} H_{N-1}^{(N-r_2+j)}
\\
-H_{N-1}^{(r_1-j)} H_{N}^{(N-r_2+j+1)}
+H_{N-1}^{(r_1-j)} H_{N}^{(N-r_2+j)}
\Big)
\end{multline}
where it is assumed that if $r\leq0$ or $r\geq N+1$ then
$H_N^{(r)}=0$ by definition.

The last expression may be equivalently expressed
in terms of the generating functions, defined
in \eqref{Hnu}, \eqref{Hnuv}:
\begin{align}\label{HNuv}
H_N(u,v)&=\frac{(u-1)H_N(u)\cdot vH_{N-1}(v) -
uH_{N-1}(u)\cdot (v-1)H_N(v)}{u-v}
\notag\\
&=\frac{1}{v-u}
\begin{vmatrix}
uH_{N-1}(u) & vH_{N-1}(v)\\
(u-1)H_{N}(u) & (v-1)H_{N}(v)
\end{vmatrix}.
\end{align}
This formula generalize to arbitrary values of the vertex weights
the result of paper \cite{S-02}
where an  equivalent expression was derived
in the case  $\lambda=\pi/2$ and $\eta=\pi/6$,
i.e., when $a=b=c$ (the so-called ice-point).

%%%%%%%%%%%%%%%%%%%%%%%%%%%%%%%%%%%%%%%%%%%%%%%%%%%%%%%%%%%%%%%%%%%
\section{A combinatorial application}

\subsection{Alternating Sign Matrices enumerations}

An alternating sign matrix (ASM)  is a matrix of $1$'s, $0$'s and $-1$'s
such that in each row and in each column \emph{(i)} all nonzero entries
alternate  in sign, and  \emph{(ii)}  the
first and the last nonzero entries  are  1. An example of
such matrix is
\begin{equation}\label{ASM}
\begin{pmatrix}
0 & 1 & 0 & 0 & 0 \\
1 & -1 & 1 & 0 & 0 \\
0 & 0 & 0 & 1 & 0 \\
0 & 1 & 0 & 0 & 0 \\
0 & 0 & 0 & 0 & 1
\end{pmatrix} .
\end{equation}
There are many nice results concerning ASMs,
for a review, see book \cite{Br-99}.
Many of these results have been first formulated as conjectures
which were subsequently proved by different methods.

The most celebrated result concerns
the total number $A(N)$ of $N\times N$ ASMs.
It was conjectured in papers \cite{MRR-82,MRR-83}
and proved in papers \cite{Z-96a,Ku-96} that
\begin{equation}\label{An}
A(N)=\prod_{k=1}^{N}\frac{(3k-2)!\, (k-1)!}{(2k-1)!\,
(2k-2)!} =
\prod_{k=1}^{N}   \frac{(3k-2)!}{(2N-k)!}.
\end{equation}

A possible generalization of the previous problem
consists in considering weighted enumerations, or
$x$-enumerations of ASMs. In the $x$-enumeration matrices are counted
with a weight $x^k$ where $k$ is the total number of
`$-1$' entries in a matrix.
The number of $x$-enumerated ASMs is denoted traditionally as
$A(N;x)$. The extension of the $x=1$ result above
to the case of generic $x$ is not known, but
for a few nontrivial cases, namely $x=2$ and $x=3$ \cite{MRR-82,MRR-83,Ku-96},
closed expressions for $x$-enumerations are known
(note that the case $x=0$ is trivial,
since assigning a vanishing weight to each
`$-1$' entry restricts enumeration to the sole permutation matrices:
$A(n;0)=n!$).

A further generalization of the previous problems  consists
in the so-called refined enumerations of ASMs, where
one counts the number of $N\times N$ ASMs
with their sole `$1$' of the last column
at the $r$-th entry. The refined enumeration can
be naturally extended to be also an $x$-enumeration.
The standard notation for the refined $x$-enumeration
is $A(N,r;x)$; in the case $x=1$ one writes simply $A(N,r)$
just like $A(N)$ for the total number of ASMs. The answer for the
refined $x$-enumeration is known only for the two values $x=1$
\cite{MRR-82,MRR-83,Z-96b} and $x=2$ \cite{MRR-83,EKLP-92}.

Such enumerations can be further generalized to the doubly refined
weighted countings, $A(N,r,s;x)$ where one counts the number
of $N\times N$ ASMs with their sole `$1$' of the first and of the last column
at the $r$-th and $(N-s+1)$-th entries, respectively.
An answer
for $A(N,r,s;x)$ in the case $x=1$ was found in \cite{S-02}.

\begin{figure}[t]
%%%%%%%%%%%%%%%%%%%%%%%%%%%%%%%%%%%%%%%%%%%%%%%%%%%%%%
%%%%%%%%%% Figure 1 %%%%%%%%%%%%%%%%%%%%%%%%%%%%%%%%%%
%%%%%%%%%%%%%%%%%%%%%%%%%%%%%%%%%%%%%%%%%%%%%%%%%%%%%%
\unitlength=1mm
\begin{center}
\begin{picture}(124,10)
% LINES
% vertex a1
\linethickness{0.5mm}
\put(7.25,10.25){\line(0,1){4.75}}
\put(7.5,10.25){\line(-1,0){5}}
\put(7.75,9.75){\line(0,-1){4.75}}
\put(7.5,9.75){\line(1,0){5}}
% vertex a2
\thinlines
\put(27,5){\line(0,1){10}}
\put(22,10){\line(1,0){10}}
% vertex b1
\linethickness{0.5mm}
\put(51.75,5){\line(0,1){10}}
\thinlines
\put(47,10){\line(1,0){10}}
% vertex b2
\thinlines
\put(72,5){\line(0,1){10}}
\linethickness{0.5mm}
\put(67,9.75){\line(1,0){10}}
% vertex c1
\linethickness{0.5mm}
\put(96.75,10){\line(0,1){5}}
\put(96.9,10){\line(-1,0){5}}
\thinlines
\put(97,9.75){\line(0,-1){5}}
\put(97,9.7){\line(1,0){4.7}}
% vertex c2
\put(117,10.25){\line(0,1){4.75}}
\put(117,10.25){\line(-1,0){5}}
\linethickness{0.5mm}
\put(117.25,10){\line(0,-1){4.75}}
\put(117,10){\line(1,0){5}}
% ASM entries
\put(6,-5){$0$}
\put(26,-5){$0$}
\put(51,-5){$0$}
\put(71,-5){$0$}
\put(96,-5){$1$}
\put(114,-5){$-1$}
\end{picture}
\end{center}
\caption{Vertex states---ASM's entries correspondence.}
\label{aentries}
\end{figure}
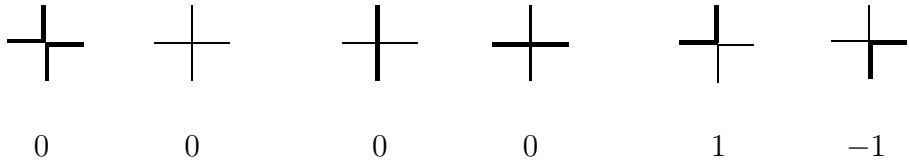

The most direct way to derive ASM enumerations is based on
the one-to-one correspondence
between $N\times N$ ASMs and configurations of the
six-vertex model on $N\times N$ lattice with DWBC,
which has been pointed out in
\cite{EKLP-92,RR-86},
and applied for the first time in \cite{Ku-96}.
The correspondence between matrix entries and vertices is depicted in
figure \ref{aentries}.
For example, matrix \eqref{ASM}
corresponds to the configuration of figure \ref{dwbcgrid} and
vice versa.

As an immediate consequence of this correspondence, ASM enumeration
is exactly given by the
partition function of square ice, when all
vertex weights are set equal to unity. More generally,
the number of `$-1$' entries in a given ASM
being equal to the number of vertices of type 6 (see figure \ref{aentries}),
and the number of vertex of type 5 and 6 being  constrained by
the condition $n_5-n_6 =N$, we readily get
\begin{equation}\label{AnZn}
A(N;x)
=(1-x/4)^{-N^2/2}\, x^{-N/2}\,
Z_N\Big|_{
\begin{subarray}{l}
\lambda=\pi/2\\
\eta=\arcsin(\sqrt{x}/2)
\end{subarray}
}.
\end{equation}
Therefore, $x$-enumeration of ASM corresponds to the computation of
the partition function of square ice on the subset of parameters space
given by $a=b$. In this correspondence,
values of $x$ belonging to the interval $(0,4)$  corresponds to the
disordered regime of the model, $-1<\Delta<1$.

This nice correspondence can be further extended to
the refined $x$-enumeration of ASMs.
In the language of square ice, the ratio  $A(N,r;x)/A(N;x)$ can
be rephrased as the probability of finding the unique vertex of type 5
on the first row at the
$(N-r+1)$-th site, which is exactly the definition  of the
boundary correlation function $H_N^{(N-r+1)}$. Explicitly, one has
\begin{equation}\label{AnrHnr}
\frac{A(N,r;x)}{A(N;x)}
=H_N^{(N-r+1)}\Big|_{
\begin{subarray}{l}
\lambda=\pi/2\\
\eta=\arcsin(\sqrt{x}/2)
\end{subarray}
}.
\end{equation}
Note that for the particular value $\lambda=\pi/2$, which is
the one of interest in ASMs enumerations, the one-point boundary correlator
enjoys the symmetry $H_N^{(r)}=H_N^{(N-r+1)}$.

Analogously, the doubly refined $x$-enumeration of ASMs is simply
related to the two-point boundary correlation function $H_N^{(r,s)}$
as follows:
\begin{equation}\label{AnrsHnrs}
\frac{A(N,r,s;x)}{A(N;x)}
=H_N^{(N-r+1,s)}\Big|_{
\begin{subarray}{l}
\lambda=\pi/2\\
\eta=\arcsin(\sqrt{x}/2)
\end{subarray}
}.
\end{equation}

It is evident from the previous formulae that  $1$-, $2$- and
$3$-enumerations of ASMs correspond to the values $\eta=\pi/6$, $\pi/4$,
$\pi/3$, respectively, in the six-vertex model.
In these three cases, the orthogonal polynomials appearing
in the representations of section \ref{sec.orthog} for the partition
function and boundary correlators turn out to specialize to so-called
classical ones, i.e. appearing into the Askey scheme of hypergeometric
orthogonal polynomials. Such polynomials can be expressed
as terminating hypergeometric series,
and hence, their characteristic properties, such as orthogonality condition,
three-term relation, etc., can be worked out explicitely (for full
details, and $q$-analog extensions, see \cite{KS-98}).
This  hypergeometric structure  allows one to evaluate
ASM enumerations in a simple and explicit way
\cite{CP-05a}.

%%%%%%%%%%%%%%%%%%%%%%%%%%%%%%%%%%%%%%%%%%%%%%%%%%%%%%%%%%%%%
\subsection{$1$-enumerations of ASMs}

To illustrate the method we shall  here restrict ourselves
to the case of $1$-enumerations of ASMs. This corresponds to
the so-called ice-point, or $\Delta=1/2$ symmetric point, of
the six-vertex model. The  parameters of the model
are specialized to $\eta=\pi/6$ and $\lambda=\pi/2$,
and correspondingly, the Boltzmann weights assume the values
$a=b=c=1$. The  orthogonality weight reads
\begin{equation}\label{mu-one}
\mu(x)=\frac{\sinh\frac{\pi}{6}x}{\sinh\frac{\pi}{2} x}
=\frac{1}{4 \pi^2}
\bigg|\Gamma\biggl(\frac{1}{3}+i\frac{x}{6}\biggr)
\Gamma\biggl(\frac{2}{3}+i\frac{x}{6}\biggr)\bigg|^2.
\end{equation}
Direct inspection of some tables of classical orthogonal polynomials, such
as \cite{KS-98}, allows to recognize here a particular specialization
of the orthogonality weight for Continuous Hahn polynomials,
which are defined as
\begin{equation}
p_n(x;a,b,c,d)
=\mathrm{i}^n
\frac{(a+c)_n(a+d)_n}{n!}
\Fthreetwo{-n}{n+a+b+c+d-1}{a+\mathrm{i} x}{a+c}{a+d}{1},
\end{equation}
and satisfy  the orthogonality relation
\begin{multline}\label{ortCHP}
\frac{1}{2\pi} \int_{-\infty}^{\infty}
p_n(x;a,b,c,d) p_m(x;a,b,c,d)\,
\Gamma(a+\mathrm{i} x)\,\Gamma(b+\mathrm{i} x)\,
\Gamma(c-\mathrm{i} x)\,\Gamma(d-\mathrm{i} x)
\,\mathrm{d}x
\\
=\frac{\Gamma(n+a+c)\,\Gamma(n+a+d)\,\Gamma(n+b+c)\,\Gamma(n+b+d)}
{(2n+a+b+c+d-1)\,\Gamma(n+a+b+c+d-1)\,n!}\, \delta_{nm}\,.
\end{multline}
Orthogonality condition \eqref{ortCHP} is valid if
the parameters $a,b,c,d$ satisfy $\Re(a,b,c,d)>0$, $a=\bar c$ and
$b=\bar d$. Comparing \eqref{mu-one} with \eqref{ortCHP} we are naturally
led to the choice of parameters $a=c=1/3$ and $b=d=2/3$. Hence,
the appropriate polynomials to be associated to the Hankel determinant
in this case are
\begin{equation}\label{poly1}
P_n(x)=
p_n\biggl(\frac{x}{6};\frac{1}{3},\frac{2}{3},
\frac{1}{3},\frac{2}{3}\biggr)
=\mathrm{i}^n (2/3)_n
\,\Fthreetwo{-n}{n+1}{1/3+\mathrm{i} x/6}{2/3}{1}{1}.
\end{equation}
The normalization constant and the leading coefficient are readily computed:
\begin{equation}%\label{}
h_n=\frac{2 (3n+1)!}{(2n+1)\, 3^{3n+1/2}\, n!},\qquad
\kappa_n=\frac{(2n)!}{6^n\, (n!)^2}
\end{equation}

Substituting the obtained values of $h_n$ and $\kappa_n$ in
expression \eqref{detZ} for the partition function, and cancelling
whatever possible, we arrive to the following value
for the ice point partition function
\begin{equation}
Z_N\Big|_{
\begin{subarray}{l}
\lambda=\pi/2\\
\eta=\pi/6
\end{subarray}
}
=\biggl(\frac{\sqrt{3}}{2}\biggr)^{N^2}\prod_{n=0}^{N-1}
\frac{(3n+1)!\, n!}{(2n)!\,(2n+1)!}
.
\end{equation}
The product expression here gives exactly the total number of ASMs,
$A(N)$, since by formula \eqref{AnZn} the first factor
relates $A(N)$ with the partition function,
\begin{equation}
A(N)
=(3/4)^{-N^2/2} \,
Z_N\Big|_{
\begin{subarray}{l}
\lambda=\pi/2\\
\eta=\pi/6
\end{subarray}
}\;.
\end{equation}
Thus, we have easily recovered the celebrated result, Eqn.~\eqref{An},
for $1$-enumeration of ASMs. It is worth noting that the
proof presented here is  considerably simpler
in comparison to those  of papers \cite{Z-96a,Ku-96}.

The refined enumeration $A(N,r)$ can be obtained within
the same framework, the derivation being only slightly more involved.
The key ingredient is now that the orthogonal polynomials belonging
to the Askey scheme are known to satisfy
some differential or finite difference equation in their variable.
In the present case, such equation reads
\begin{multline}\label{findiffeq}
\left(\frac{1}{3}-\frac{\rmi x}{6}\right)
\left(\frac{2}{3}-\frac{\rmi x}{6}\right) P_{N-1}(x+6\rmi)
+\left[\frac{x^2}{18}-\frac{4}{9}- N(N-1)\right] P_{N-1}(x)
\\
+\left(\frac{1}{3}+\frac{\rmi x}{6}\right)
\left(\frac{2}{3}+\frac{\rmi x}{6}\right)
P_{N-1}(x-6\rmi)
=0\,.
\end{multline}

First we recall that the refined enumeration is, modulo an
obvious overall normalization factor, see \eqref{AnrHnr}, nothing but
the one-point boundary correlator, which in the present case
($\eta=\pi/6$ and $\lambda=\pi/2$) reads:
\begin{equation}\label{bulkcorrbis}
H_N^{(r)}= H_N^{(N-r+1)}=\const \Bigl\{
P_{N-1}(\partial_\eps)
[\varrho(\eps)]^{N-1} [\omega(\epsilon)]^{r-1}\Bigr\}\Big|_{\eps=0}
\end{equation}
where $\omega(\eps)$ and $\varrho(\eps)$ are given by
\begin{equation}\label{og1}
\omega(\epsilon)=\frac{\sin\eps}{\sin(\eps-\pi/3)},\qquad
\varrho(\epsilon)=\frac{\sin(\eps-\pi/3)}{\sin(\eps+\pi/3)}.
\end{equation}
Then,  exploiting the simple relation
$P_{N-1}(\partial_\eps\pm 6\rmi)=\rme^{\mp 6\rmi\eps}
P_{N-1}(\partial_\eps) \rme^{\pm 6\rmi\eps}$, it is easy to see
that finite difference
equation \eqref{findiffeq} implies the condition
\begin{equation}\label{bulkcorr1}
P_{N-1}(\partial_\eps)\
\left[ \sin3\eps\,\partial_\eps^2\,\sin3\eps
+\sin^2 3\eps-9 N(N-1)\right]\
[\varrho(\epsilon)]^{N-1} [\omega(\epsilon)]^{r-1} \tau(\epsilon)
\Big|_{\eps=0}=0
\end{equation}
where $\tau(\eps)$ is at this stage an  arbitrary function.
Our aim now is to determine the form of $\tau(\eps)$
in such a way that the last equation can be turned into a recurrence
relation in $r$ for the boundary correlator $H_N^{(r)}$.
A constructive procedure has been devised in \cite{CP-05a} to
perform this task in general. The derivation is given there in full detail.
In the present case the resulting
recurrence relation reads
\begin{equation}\label{recur1}
r(r-2N+1) H_N^{(r+1)}
-(r-N)(N+r-1)H_N^{(r)}=0.
\end{equation}
This recurrence  can be easily solved modulo a
normalization constant
\begin{equation}
H_N^{(r)}= \const  \frac{(N+r-2)!\,(2N-1-r)!}{(r-1)!\,(N-r)!}.
\end{equation}

A possible way to fit the normalization condition \eqref{normcond},
is to consider the generating function
$H_N(z)$ defined via Eqn.~\eqref{Hnu}. The result reads
\begin{align}\label{Hnz1}
H_N(z)= \frac{(2N-1)!\,(2N-2)!}{(N-1)!\,(3N-2)!}\;\F{1-N}{N}{2-2N}{z}
\end{align}
where the proper normalization is easily determined
through Chu-Vandermonde identity
\begin{equation}\label{Gauss-sum}
\F{-m}{b}{c}{1}
=\frac{(c-b)_m}{(c)_m};\qquad
(a)_m:=a(a+1)\cdots(a+m-1).
\end{equation}\
Inspecting the coefficient of $z^{r-1}$ in \eqref{Hnz1}
one obtains
\begin{equation}\label{Hnr1}
H_N^{(r)}=\frac{\binom{N+r-2}{N-1}\binom{2N-1-r}{N-1}}{\binom{3N-2}{N-1}}.
\end{equation}
The refined $1$-enumeration of ASMs, conjectured in \cite{MRR-83},
and first proven in \cite{Z-96b},
immediately follows from relation \eqref{AnrHnr}.

Let us conclude with the doubly refined $1$-enumeration of ASMs,
which is readily obtained by substituting expression \eqref{Hnr1}
for the refined enumeration into \eqref{H=HH}.
Analogously, the corresponding generating function, $H_N(u,v)$,
is immediately obtained by substituting \eqref{Hnz1} into \eqref{HNuv},
hence recovering the result of paper \cite{S-02} where a different
approach was considered.

\subsection{Other ASMs enumerations}

We have just seen that the present approach allows a very
simple, straightforward and unified derivation of all
answers  concerning $1$-enumerations of ASMs.
These answers were already known,
but the method exposed above can be extended to all $x$-enumerations
where the underlying orthogonal polynomials appear to belong
to the Askey scheme. Indeed, when this is the case, $h_n$, $\kappa_n$
are explicitly known, and moreover the considered orthogonal polynomials
happen to satisfy a finite  difference equation in their variable;
has illustrated above, this is sufficient to compute explicitly
the ordinary, refined, and doubly refined enumerations.
As a matter of fact, it appears that the underlying orthogonal polynomials
belong to the Askey scheme only for $x=1,2,3$ (for completeness we
mention also the trivial case $x=0$, where Meixner-Pollaczek
orthogonality measure naturally emerges when the $\eta=0$ limit is performed
in \eqref{meas}, and the   highly non-trivial case $x=4$, where
the corresponding orthogonal polynomials turns out to be of Bannai-Ito type
\cite{BI-84}, and therefore do not fall within Askey scheme).
Our approach allows to recover in a straightforward way the results
concerning $2$-enumerations, and  to derive the previously unknown
answers for the refined and doubly refined $3$-enumerations.

The $2$-enumerations correspond to the so-called free-fermion point
of the square ice, where the parameters of the model
assume the values $\lambda=\pi/2$, $\eta=\pi/4$. As a matter of fact,
the whole free-fermion line of the model ($\Delta=0$) can be treated at once,
with the crossing parameter set to the value $\eta=\pi/4$, while
$\lambda$ is free to vary in the interval $\pi/4<\lambda<3\pi/4$.
In this case the associated orthogonal polynomials can be recognized
as a particular specialization of Meixner-Pollaczek
polynomials \cite{KS-98}.
The computation of the partition function
and boundary  correlators, and thus of the various ASM $2$-enumerations,
can  be performed  along the lines of the
proposed approach \cite{CP-05a}, and is particularly simple.
The various answers in this case have been known for some time
\cite{MRR-83,ICK-92,EKLP-92,BPZ-02}.

Let us now discuss the $3$-enumerations,
which correspond to the so-called dual ice-point ($\Delta=-1/2$)
of the six-vertex model, with $\lambda=\pi/2$ and $\eta=\pi/3$.
In this case one can recognize that the measure is associated to
the Continuous Dual Hahn polynomials \cite{KS-98}. A
complication however occurs: on one hand these are polynomials
in $x^2$ rather than in $x$, and on the other hand the support
of the  integration measure is restricted to the positive half-axis.
These inconveniences are however easily circumvented, provided that
the cases of even and odd $N$ ($N$ being the size of the lattice,
or of the ASMs) are treated separately.
Indeed the set of polynomials $P_n(x)$ associated to
the determinant representation for the partition function or  the boundary
correlation functions should be specified differently for even or odd $n$,
each of these two cases corresponding to a slightly different
choice of parameters in the Continuous Dual Hahn polynomials.
This is in fact a merely technical complication and the general
procedure previously outlined may again be applied without modification.
This has been done in full detail in \cite{CP-05a}. We shall
here just recall the main results.
The known answer for the partition function,  or, equivalently,
for the ordinary $3$-enumeration of ASMs \cite{MRR-83,Ku-96},
is readily recovered. Moreover the previously unknown answer
for the one-point boundary correlator (or, equivalently,
for the refined  $3$-enumeration of ASMs)
can be worked out, and reads:
\begin{equation}
\begin{split}
H_{2m+2}^{(r)}&=\frac{B(m,r-1)+B(m,r-2)}{2}
\\
H_{2m+3}^{(r)}&=\frac{2\,B(m,r-1)+5\,B(m,r-2)+2\,B(m,r-3)}{9}
\end{split}
\end{equation}
where the quantities $B(m, n)$,
obeying $B(m, n)=B(m,2m- n)$,
are given by
\begin{multline} \label{theor2}
B(m, n)=\frac{(2m+1)!\,m!}{3^m\,(3m+2)!}
\sum_{\ell=\max(0, n-m)}^{[ n/2]}
(2m+2- n+2\ell)\binom{3m+3}{ n-2\ell}
\\ \times
\binom{2m+\ell- n+1}{m+1} \binom{m+\ell+1}{m+1}2^{ n-2\ell}
\end{multline}
for $ n=0,1,\dots,2m$, while they are assumed to vanish otherwise.
It should be mentioned that this result has first been obtained in
\cite{CP-04} through the solution of a second order differential
equation associated to the  generating function $H_N(u)$,
see \eqref{Hnu}. Such differential equation, which on one hand is
equivalent to the recurrence relation which can be derived from
the orthogonal polynomial approach (i.e. the analogue of \eqref{recur1} for
the present case), is on the other hand closely related
to the functional equation proposed in  \cite{S-03}, within a
different approach to the problem.

The two-point boundary correlator is immediately obtained simply
by inserting the previous expression into \eqref{H=HH}.

%%%%%%%%%%%%%%%%%%%%%%%%%%%%%%%%%%%%%%%%%%%%%%%%%%%%%%%%%%%%%%%%%
\section{An alternative representation for the (partially)
inhomogeneous partition function}

We have till now restricted ourselves to the homogeneous
version of the six-vertex model. This has allowed us to apply very standard
tools from the theory of orthogonal polynomials to derive in
a simplified and unified way several nontrivial results.
Of course, in doing so, we have lost the dependence in, let us say,
the partition function,  on the spectral parameters of the model.
This dependence implicitly encodes information
about the correlation functions of the model, and is thus of great interest.
We want here to show how the results of section
\ref{sec.orthog} for  the boundary correlators, based
on the proposed orthogonal polynomial representation, can be used
to recover the dependence of the partition function on the
spectral parameters of the (partially) inhomogeneous version of
the six-vertex model.

As briefly explained in section \ref{sec.6vm},  when considering the
inhomogeneous version of the six-vertex model, one associates a spectral
parameter $\lambda_{\alpha}$ ($\alpha=1,\dots,N$) to each row of the lattice,
and $\nu_{\beta}$ ($\beta=1,\dots,N$) to each  column. We want here to consider
a particular situation in which all horizontal  inhomogeneities $\nu_{\beta}$
are set to zero, and a subset of the vertical one is set to some value
$\lambda$. For simplicity, and comparison with previous works
\cite{S-02,DfZj-04} we shall fix this value to  $\lambda=\pi/2$ (note
however that our results can be readily extended to the case of
generic values of $\lambda$). On the other hand $\eta$ will be considered
generic.
We reparametrize the Boltzmann weights \eqref{sin} as follows:
\begin{equation}\label{newparam}
a(z)=zq+1\,,\qquad b(z)=z+q\,,\qquad c(z)=(1-q^2)\sqrt{-\frac{z}{q}}\,,
\end{equation}
where
\begin{equation}
z_\alpha=
\rme^{2 i (\lambda_{\alpha} -\pi/2)}\,,
\qquad
q=\rme^{2 i \eta}\,.
\end{equation}
We have here ignored the parameters $\nu_{\alpha}$, which have all been set
to zero; the definition of the Boltzmann  weights in  \eqref{newparam}
moreover differs from \eqref{sin} by a common overall factor. This choice
turns out to be more convenient for what follows, and  can be
absorbed in the normalization of the partition function, which is
now a function of $N$ variables: $Z_N(z_1,\dots,z_N)$.

In the case when only one inhomogeneity is present, in the first column,
the partition function can be expressed in terms of the boundary correlator
$H_N^{(r)}$ as follows:
\begin{align}\label{ZnHn}
Z_N(z,1,\dots,1)&=Z_N(1,1,\dots,1)\, \sum_{r=1}^{N}
\left(\frac{b(z)}{b(1)}\right)^{r-1}
\frac{c(z)}{c(1)}
\left(\frac{a(z)}{a(1)}\right)^{N-r}
H_N^{(r)} \notag\\
&=Z_N(1,1,\dots,1)\, \frac{c(z)}{c(1)}
\left(\frac{a(z)}{a(1)}\right)^{N-1}
H_N \big(b(z)/a(z)\big)\,,
\end{align}
where the generating function $H_N(w)$, see  \eqref{Hnu}, is a polynomial
of order $N-1$ in $w$.
Let us now introduce the variable $u$, related to $z$ as follows
\begin{equation}
u=\frac{b(z)}{a(z)}=\frac{z+q}{1+zq}\,,\qquad z=\frac{u-q}{1-uq}\,,
\end{equation}
and consider the partition function as a function of the variables
$u_1,\dots, u_n$,
by introducing
\begin{equation}
{\tilde{Z}}_N(u_1,\dots,u_N)=\frac{Z_N(z_1,\dots,z_N)}{Z_N(1,\dots,1)}
\prod_{\alpha=1}^{N} \left[\frac{c(1)}{c(z_{\alpha})}
\left(\frac{a(1)}{a(z_{\alpha})}\right)^{N-1}\right]\,.
\end{equation}
Note that the `partition function' $\tilde{Z}_N(u_1,\dots,u_N)$ is
normalized such that $\tilde{Z}_N(1,\dots,1)=1$.
Relation \eqref{ZnHn} then  reads
\begin{equation}
{\tilde{Z}}_N\left(u,1,\dots,1\right)=H_N(u)\,.
\end{equation}
In the same way, following \cite{S-02,DfZj-04}, we may rewrite
the partition function with two inhomogeneities as
the generating  function \eqref{Hnuv} of the two-point
boundary correlator. Since the latter is in turn expressible in terms
one-point boundary correlators, see \eqref{HNuv}, we immediately  have
\begin{equation}
{\tilde{Z}}_N\left(u_1,u_2,1,
\dots,1\right)=\frac{1}{\varDelta(u_1,u_2)}
\begin{vmatrix}
u_1 H_{N-1}(u_1) & u_2 H_{N-1}(u_2)\\
(u_1-1)H_{N}(u_1) & (u_2-1)H_{N}(u_2)
\end{vmatrix}.
\end{equation}
It is now quite natural to guess that
\begin{multline}\label{Zinhom}
{\tilde{Z}}_N\left(u_1,\dots, u_k ,1,
\dots,1\right)
=\frac{1}{\varDelta(u_1,\dots,u_k)}\\
\times
\begin{vmatrix}
u_1^{k-1} H_{N-k+1}(u_1) & u_2 ^{k-1}H_{N-k+1}(u_2)
&\dots&u_k^{k-1}H_{N-k+1}(u_k)\\
u_1^{k-2} (u_1 -1) H_{N-k+2}(u_1) & u_2^{k-2}(u_2 -1) H_{N-k+2}(u_2)
&\dots&u_k^{k-2} (u_k -1)H_{N-k+2}(u_k)\\
\hdotsfor{4}  \\
(u_1-1)^{k-1}H_{N}(u_1) & (u_2-1)^{k-1}H_{N}(u_2)
&\dots& (u_k -1)^{k-1}H_{N}(u_k)
\end{vmatrix}.
\end{multline}
We note that the right hand side is by construction a symmetric
polynomial in $u_1, \dots, u_k$, of order $N-1$ in each variable,
as it should. Moreover it is evident that
the homogeneous limit can be performed one variable at a time, say
$u_k\to 1$, maintaining the proposed structure: formula \eqref{Zinhom}
obviously satisfies
\begin{equation}
\lim_{u_k\to 1}
{\tilde{Z}}_N\left(u_1,\dots,
u_{k-1}, u_k,1,\dots,1\right)=
{\tilde{Z}}_N\left(u_1, \dots,
u_{k-1},1,1,\dots,1\right)\,.
\end{equation}
It is also straightforward to verify  that
\begin{equation}
\lim_{u_k\to 0}
{\tilde{Z}}_N\left(u_1,\dots,
u_{k-1},u_k,1,\dots,1\right)=
H_N(0)\,
{\tilde{Z}}_{N-1}\left(u_1,\dots,
u_{k-1},1,\dots,1\right)\,,
\end{equation}
which is nothing but Korepin recursion relation \cite{K-82}
specialized to the present situation. It should however
be mentioned that in the
present case, where all `horizontal' spectral parameters $\nu_{\beta}$
have been set to same value, the uniqueness of the solution of such
a recursion relation is not guaranteed.
A complete derivation of representation \eqref{Zinhom} can however
be given by exploiting a set of identities relating the derivatives of
function $\varphi(\lambda,\eta)$ to the  polynomials  $H_N(u)$; this will
be done elsewhere.

We would like to conclude by emphasizing that representation \eqref{Zinhom}
constitutes a rather wide generalization of previous expressions
given in \cite{S-02,DfZj-04}. First of all it generalizes the previous
formulae, holding in the case of two spectral parameters,
to a larger number of variables.
Moreover, the proposed representation holds for any values of $\eta$,
(and can be extended straightforwardly to any value of $\lambda\not=\pi/2$,
by a suitable redefinition of variable $u$),
expressing in general the partition function in terms of
just one-point boundary  correlators of the corresponding homogeneous model.

%%%%%%%%%%%%%%%%%%%%%%%%%%%%%%%%%%%%%%%%%%%%%%%%%%%%%%%%%%%%%%%%%%%
\section*{Acknowledgments}

One of us (FC) is grateful to the Centre de Recherches Math\'ematiques \`a
Montr\'eal, and to all those who contributed to the realization
of the short  program on `Random Matrices, Random Processes and Integrable
Systems'. A special thank to John Harnad,  Jacques Hurtubise and Marco
Bertola for the warm atmosphere and the pleasant stay.
We acknowledge financial support from MIUR PRIN programme (SINTESI
2004). One of us (AGP) is also supported in part by Civilian Research
and Development Foundation (CRDF grant RUM1-2622-ST-04), by
Russian Foundation for Basic Research (RFFI grant
04-01-00825), and by the programme Mathematical Methods in Nonlinear
Dynamics of Russian Academy of Sciences. This work is partially
done within the European Community network EUCLID (HPRN-CT-2002-00325).

%%%%%%%%%%%%%%%%%%%%%%%%%%%%%%%%%%%%%%%%%%%%%%%%%%%%%%%%%%%%%%%%%%%
%%%%%%%%%%%%%%%%%%%%%%%%%%%%%%%%%%%%%%%%%%%%%%%%%%%%%%%%%%%%%%%%%%%


\begin{thebibliography}{**}
%\nospacing\small

\bibitem{K-82}
V.E. Korepin.
Calculations of norms of Bethe wave functions.
\textit{Commun. Math. Phys.} (1982) \textbf{86} 391--418.

\bibitem{KBI-93}
V.E. Korepin, N.M. Bogoliubov, and A.G. Izergin.
\textit{Quantum Inverse Scattering Method and Correlation Functions}.
Cambridge University Press, Cambridge, 1993.

\bibitem{I-87}
A.G. Izergin. Partition function of the six-vertex model in
the finite volume.
\textit{Sov. Phys. Dokl.} \textbf{32} (1987) 878--879.

\bibitem{ICK-92}
A.G. Izergin, D.A. Coker, and V.E. Korepin.
Determinant formula for the six-vertex model.
\textit{J. Phys. A: Math. Gen.} \textbf{25} (1992) 4315--4334.

\bibitem{BPZ-02}
N.M. Bogoliubov, A.G. Pronko, and M.B. Zvonarev.
Boundary correlation functions of the six-vertex model.
\textit{J. Phys. A: Math. Gen.} \textbf{35} (2002) 5525--5541.

\bibitem{CP-05b}
F. Colomo and A.G. Pronko.
On two-point boundary correlations in the six-vertex model with DWBC.
\textit{J. Stat. Mech.: Theor. Exp.} JSTAT(2005)P05010.
e-Print, arXiv:  math-ph/0503049.

\bibitem{LW-72}
E.H. Lieb and F.Y. Wu.
In: \textit{Phase Transitions and Critical Phenomena}, Vol. 1,
edited by C. Domb and M. S. Green, Academic Press, London, 1972,
p.~321.

\bibitem{B-82}
R.J. Baxter.
\textit{Exactly Solved Models in Statistical Mechanics}.
Academic press, San Diego, 1982.

\bibitem{Br-99}
D.M. Bressoud.
\textit{Proofs and Confirmations: The Story of the
Alternating Sign Matrix Conjecture}.
Cambridge University Press, Cambridge, 1999.

\bibitem{RS-01}
A.V. Razumov and  Yu.G. Stroganov.
Spin chains and combinatorics.
\textit{J. Phys. A: Math. Gen.} \textbf{34} (2001) 3185--3190.

\bibitem{RS-04}
A.V. Razumov and  Yu.G. Stroganov.
Combinatorial nature of ground state vector of O(1) loop model.
\textit{Theor.Math.Phys.} \textbf{138} (2004) 333-337;
\textit{Teor.Mat.Fiz.} \textbf{138} (2004) 395-400.
e-Print, arXiv: math.CO/0104216.

\bibitem{NRdG-05}
A. Nichols, V. Rittenberg, J. de Gier.
One-boundary Temperley-Lieb algebras in the XXZ and loop models.
\textit{J. Stat. Mech: Theor. Exp.} JSTAT(2005)P03003.
e-Print, arXiv: cond-mat/0411512.

\bibitem{dGN-05}
J. de Gier and B. Nienhuis.
Brauer loops and the commuting variety.
\textit{J. Stat. Mech: Theor. Exp.} JSTAT(2005)P01006.

\bibitem{DfZj-05}
P. Di Francesco and P. Zinn-Justin.
Quantum Knizhnik-Zamolodchikov equation, generalized Razumov-Stroganov
sum rules and extended Joseph polynomials.
e-Print, arXiv: math-ph/0508059.

\bibitem{Ku-96}
G. Kuperberg.
Another proof of the alternative-sign matrix conjecture.
\textit{Internat. Math. Res. Notices} \textbf{1996} (1996) 139--150.

\bibitem{Z-96b}
D. Zeilberger.
Proof of the refined alternating sign matrix conjecture.
\textit{New York J. Math} \textbf{2} (1996) 59--68.

\bibitem{S-02}
Yu. G. Stroganov.
A new way to deal with Izergin--Korepin determinant at root of unity.
e-Print, arXiv: math-ph/0204042.


\bibitem{DfZj-04}
P. Di Francesco and P. Zinn-Justin.
Around the Razumov-Stroganov conjecture: proof of a multi-parameter
sum rule.
e-Print, arXiv: math-ph/0410061.

\bibitem{KS-98}
R. Koekoek and R.F. Swarttouw.
The Askey-scheme of hypergeometric orthogonal polynomials and its
$q$-analoque. Report no. 98-17, Delft University of Technology, 1998.

\bibitem{dGK-01}
J. de Gier and V. Korepin.
Six-vertex model with domain wall boundary conditions. Variable
inhomogeneities.
\textit{J. Phys. A: Math. Gen.} \textbf{34} (2001) 8135--8144.
e-Print, arXiv: math-ph/0101036.

\bibitem{KZj-00}
V.E. Korepin and P. Zinn-Justin.
Thermodynamic limit of the six-vertex model
with domain wall boundary conditions.
\textit{J. Phys. A: Math. Gen.} \textbf{33} (2000) 7053--7066.

\bibitem{Zj-00}
P. Zinn-Justin.
Six-vertex model with domain wall boundary conditions
and one-matrix model.
\textit{Phys. Rev. E} \textbf{62} (2000) 3411--3418.

\bibitem{CP-05a}
F. Colomo and A.G. Pronko.
Square ice, alternating sign matrices, and classical orthogonal
polynomials.
\textit{J. Stat. Mech.: Theor. Exp.} JSTAT(2005)P01005.
e-Print, arXiv:  math-ph/0411076.

\bibitem{MRR-82}
W.H. Mills, D.P. Robbins, and H. Rumsey.
Proof of the Macdonald conjecture.
\textit{Invent. Math.} \textbf{66} (1982) 73--87.

\bibitem{MRR-83}
W.H. Mills, D.P. Robbins, and H. Rumsey.
Alternating-sign matrices and descending plane partitions.
\textit{J. Combin. Theory Ser. A} \textbf{34} (1983) 340--359.

\bibitem{Z-96a}
D. Zeilberger.
Proof of the alternating sign matrix conjecture.
\textit{Elec. J. Comb.} \textbf{3} (2) (1996) R13.

\bibitem{EKLP-92}
N. Elkies, G. Kuperberg, M. Larsen, and J. Propp.
Alternating-sign matrices and domino tilings.
\textit{J. Algebraic Combin.} {\bf 1} (1992) 111--132;
\textit{ibid.} 219--234.

\bibitem{RR-86}
D.P. Robbins and H. Rumsey.
Determinants and alternating-sign matrices.
\textit{Advances in Math.} \textbf{62} (1986) 169--184.

\bibitem{S-75}
G. Szeg\"o.
\textit{Orthogonal Polynomials}.
Fourth edition, Colloquium Publications,
Vol. 23, Amer. Math. Soc., Providence, RI, 1975.

\bibitem{BI-84}
E. Bannai and T. Ito.
\textit{Algebraic combinatorics. I. Association schemes.}
The Benjamin/Cummings Publishing Co., Inc., Menlo Park, CA, 1984.

\bibitem{CP-04}
F. Colomo and A.G. Pronko.
On the refined 3-enumeration of alternating sign matrices.
\textit{Advances in Applied Mathematics} \textbf{34} (2005) 798--811.
e-Print, arXiv:  math-ph/0404045.

\bibitem{S-03}
Yu. G. Stroganov.
3-enumerated alternating sign matrices.
e-Print, arXiv: math-ph/0304004.

\end{thebibliography}
\end{document}